\documentclass[useAMS,usenatbib]{mn2e}
\usepackage{graphicx,epsfig}

\def\eg{{e.g.,~}}
\def\etal{{et\,al. }}
\def\asec{\ifmmode ^{\prime\prime}\else$^{\prime\prime}$\fi}

\def\msun{\hbox{~M$_{\odot}$}}

\def\degs{\ifmmode ^{\circ}\else$^{\circ}$\fi}
\def\amin{\ifmmode ^{\prime}\else$^{\prime}$\fi}
\def\asec{\ifmmode ^{\prime\prime}\else$^{\prime\prime}$\fi}
\def\farcs{\hbox{$.\!\!^{\prime\prime}$}}  
\def\h{\hbox{$^{\rm h}$}}
\def\m{\hbox{$^{\rm m}$}}

\def\degs{\ifmmode ^{\circ}\else$^{\circ}$\fi}
\def\amin{\ifmmode ^{\prime}\else$^{\prime}$\fi}

\def\EE#1{\times 10^{#1}}

\def\cm{\mbox{\,cm}}

\def\cm3{\mbox{\,cm$^{-3}$}}
\def\kms{\mbox{\,km~s$^{-1}$}}

\def\erg{\mbox{\,erg}}
\def\ergs{\mbox{\,erg~s$^{-1}$}}

\def\kms{\mbox{\,km s$^{-1}$}}

\def\lsim{\!\!\!\phantom{\le}\smash{\buildrel{}\over
 {\lower2.5dd\hbox{$\buildrel{\lower2dd\hbox{$\displaystyle<$}}\over
                                 \sim$}}}\,\,}
\def\gsim{\!\!\!\phantom{\ge}\smash{\buildrel{}\over
{\lower2.5dd\hbox{$\buildrel{\lower2dd\hbox{$\displaystyle>$}}\over
                               \sim$}}}\,\,}

\def\Lya{{\rm\,Ly-$\alpha$~}}
\newcommand\HeII{He{\small II}~$\lambda$\,1640}
\newcommand\Hbeta{H$\beta$}
\newcommand\OII{[O{\small II}]~$\lambda$\,3727}
\newcommand\OIII{[O{\small III}]~$\lambda$\,5007}

\title{The extremely asymmetric radio structure of the z=3.1 radio galaxy B3~J2330+3927}

\author[M.~A P\'erez-Torres and C. De Breuck] {M-A. P\'erez-Torres$^1$
       and C. De Breuck$^{2}$ \\ 
       $^1$Instituto de Astrof\'{\i}sica de
       Andaluc\'{\i}a, CSIC, Apdo. Correos 3004, E-18080 Granada,
       Spain\\ 
       $^2$European Southern Observatory, Karl Schwarzschild
       Stra\ss e 2, D-85748 Garching, Germany\\ }

\date{Accepted  2005 July 15.
      Received  2005 July 14;
      in original form 2005 June 29.
     }

\pagerange{\pageref{firstpage}--\pageref{lastpage}} \pubyear{2005}
\def\LaTeX{L\kern-.36em\raise.3ex\hbox{a}\kern-.15em
T\kern-.1667em\lower.7ex\hbox{E}\kern-.125emX}

\begin{document}
\label{firstpage}
\maketitle

\begin{abstract}

We report on 1.7 and 5.0 GHz observations of the $z=3.087$ radio
galaxy B3~J2330+3927, using the Very Long Baseline Array (VLBA), and
archival 1.4 and 8.4~GHz Very Large Array (VLA) data.  Our VLBA data
identify a compact, flat spectrum ($\alpha_{\rm 1.7~GHz}^{\rm 5~GHz} =
-0.2 \pm 0.1; S_\nu \propto \nu^{\alpha}$) radio component as the
core.  The VLA images show that the fraction of core emission is very
large ($f_c\approx 0.5$ at 8.4~GHz), and reveal a previously
undetected, very faint counterjet, implying a radio lobe flux density
ratio $R\gsim$11 and a radio lobe distance ratio $Q\approx$1.9.  Those
values are much more common in quasars than in radio galaxies, but the
optical/near-IR spectra show a clear type~II AGN for B3~J2330+3927,
confirming that it is indeed a radio galaxy.  Unlike all other radio
galaxies, the bright \Lya emitting gas is located towards the furthest
radio arm.  We argue against environmental and relativistic beaming
effects being the cause of the observed asymmetry, and suggest this
source has intrinsically asymmetric radio jets. If this is the case,
B3~J2330+3927 is the first example of such a source at high redshift,
and seems to be difficult to reconcile with the unified model, which
explains the differences between quasars and radio galaxies as being
due to orientation effects.

\end{abstract}


\section{Introduction}
\label{intro}

High redshift radio galaxies (HzRGs, $z>2$) are excellent probes to
explore massive galaxies and their proto-cluster environments in the
early Universe \citep[\eg][]{debreuck02,rocca04,venemans04}. HzRGs are
also ideal tools for probing the activity of AGN in their host
galaxies.  Indeed, the powerful jets and ionizing radiation from young
radio AGN can dramatically affect the environment of their forming
host galaxies.  This is clearly illustrated by the ``alignment
effect'' of optical line-emission and UV-continuum along the radio
jets (\eg \citealt{chambers87}; \citealt{mccarthy91}), and recent
discoveries of similarly radio/X-ray aligned emission (\eg\
\citealt{carilli02}, \citealt{scharf03}, \citealt{overzier05}).
Collimated radiation from the AGN may scatter off surrounding
material, while the jets shock and heat the ambient gas and may even
trigger star formation (\eg\ \citealt{bicknell00}).  Thus HzRGs may
affect, at least temporarily, the evolution of their hosts.

High-resolution radio imaging of HzRGs, especially if combined with
high-resolution optical and X-ray studies, is essential to
make progress on some of the above issues. In particular,
morphological comparisons of radio and optical continuum images may
help identify effects of jet-induced star formation.  Similarly,
comparisons of high-resolution radio and \Lya\ images may point to the
interaction between the propagating jet and the surrounding primeval
interstellar medium. The importance of such environmental effects is
well established by the correlated radio and optical asymmetries in
powerful radio sources \citep{mccarthy91}.

Radio galaxies at high redshifts are inevitably younger, and
have smaller angular sizes than their lower redshift counterparts
(\citealt{blundell99}). In many cases, their central structures are
much smaller than 1\arcsec, and cannot be resolved even with
the VLA in A-array. Very-long-baseline interferometry (VLBI) is
therefore needed to resolve the central components and identify their
radio cores. The milliarcsecond (mas) resolution obtained with VLBI 
also allows us to
morphologically identify the radio cores, and to study the radio jet
structures.

In this {\it Letter}, we present high-resolution imaging of the HzRG
B3~J2330+3927 (4$-$11~mas resolution), obtained with the VLBA at 1.7
and 5~GHz.  We also present archival 1.4 and 8.4~GHz data from the
VLA.  Our VLBA observations clearly identify the radio core with a
compact, flat spectrum source, coincident with the optical/near-IR AGN
emission and the peak of the CO emission. The deep 1.4~GHz VLA image
shows a very weak counterjet, implying that B3~J2330+3927 has one of
the most asymmetric radio structures ever reported for a type~II AGN.
We assume a $\Lambda$-dominated universe with $\Omega_{\rm M}=0.3$,
$\Omega_\Lambda = 0.7$, and $H_0 = 65\kms$~Mpc$^{-1}$.  At $z=3.087$,
1\arcsec\ corresponds to 8.2~kpc.

\section{The high-redshift radio galaxy B3~J2330+3927}

The radio galaxy B3~J2330+3927 has been comprehensively studied at
several frequencies by \citet{debreuck03}, hereafter DB03.  The
8.4~GHz VLA observations of DB03 showed that B3~J2330+3927 is a $\sim
2\asec$ wide source consisting of three radio components, located in
between two optical ($K-$band) objects ({\em a} and {\em b}; see
Fig.~2 of DB03).  Based on its radio morphology, DB03 suggested that
the marginally resolved central component might be the radio core.
However, optical and near-IR Keck spectroscopy of object {\em a}
showed that B3~J2330+3927 has a classical AGN type II spectrum at
$z=3.087$, suggesting that this object hosts the radio core.  If this
is the case, a very peculiar one-sided jet morphology was then implied for
B3~J2330+3927.

The 113~GHz observations of DB03 with the IRAM Plateau de Bure
Interferometer revealed CO~{J=4-3} emission peaking at a position
coincident with component {\em a}, and at a velocity that closely
corresponded to the velocity shift of an associated H~I absorber seen
in \Lya. This strongly suggested that both CO emission and H~I
absorption originated from the same gas reservoir surrounding the host
galaxy.  
DB03 estimated a molecular hydrogen mass M(H$_2$) = 7$\times10^{10}\msun$, indicating that the host galaxy must be surrounded by a massive gas reservoir.

\begin{figure}
\vspace{-0.4cm}
\begin{center}
\includegraphics[width=8.5cm]{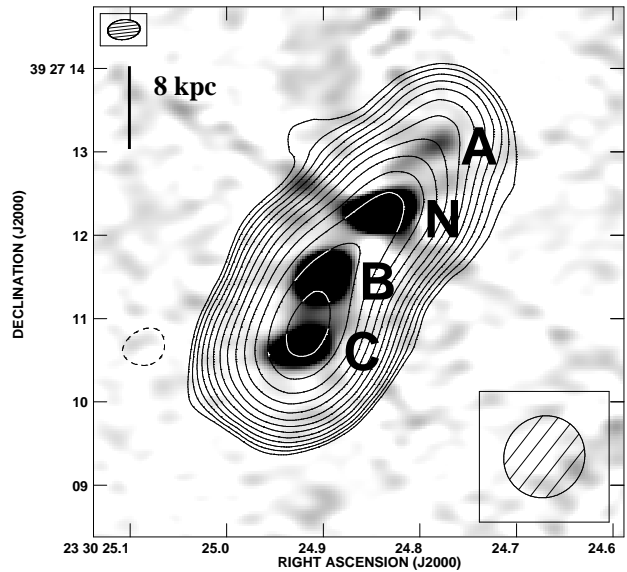}
\caption{VLA A-array images of B3~J2330+3927. Greyscales show the
naturally weighted 8.4\,GHz image, and contours the uniformly weighted
1.4\,GHz image. The contour scheme is a geometric progression in
$\sqrt{2}$.
The first contour level is at 0.6\,mJy beam$^{-1}$.  The synthesized
beams are shown in the top left (8.4\,GHz) and bottom right (1.4\,GHz)
corners. Note the confirmation of the weak 8.4\,GHz component A by the
1.4\,GHz contours.  }
\label{fig,vla}
\end{center}
\end{figure}

\section{Radio observations of B3~J2330+3927}
\label{sec,radio}

\subsection{VLA observations}
\label{sec,vla}

We reanalyzed the 8.4~GHz observations of B3~J2330+3927 (DB03) made on
 2002 March 30 with the VLA in A-configuration.  The observations were
made in standard continuum mode, 
using a bandwidth of 50~MHz.
We used a scan on 3C~48 from the previous observations
to set the absolute VLA flux density scale.  The compact source J2257+4154
was used as the phase calibrator. We used standard 
amplitude calibration and phase self-calibration
techniques within {\it AIPS}
 to obtain the 8.4~GHz VLA
image of B3~J2330+3927 in Fig.~\ref{fig,vla}. 
The total cleaned flux density in the image is $\approx$21.4~mJy and 
the image off-source rms $\sim$40$\mu$Jy/beam.

We also analyzed archival 1.4~GHz data obtained with the VLA in
A-configuration on 20 January 2001, kindly provided by R. Ivison.  The
observations were done in pseudo-continuum, spectral line mode with
7$\times$3.125\,MHz bandwidth and a central frequency of 1.400\,GHz.
We imaged B3~J2330+3927 using uniform weighting, resulting in
a synthesized beam size of 1\farcs00 $\times$ 0\farcs95. 
The total cleaned flux density in the image is $\approx$104.3~mJy and 
the image off-source rms $\sim$60$\mu$Jy/beam.

\subsection{VLBA observations}
\label{sec,vlba}

We observed B3~J2330+3927 on 2004 November 29 and December 9 with the
VLBA at 1.7 and 5~GHz, respectively.  We observed at a bit rate of 128
Mbit/s, and each frequency band was split into four intermediate
frequencies (IFs) of 8 MHz bandwidth each, for a total synthesized
bandwidth of 32~MHz .  Each IF was in turn split into 32 channels of
0.25~MHz bandwidth each.  The data were correlated at the VLBA
Correlator of the National Radio Astronomy Observatory (NRAO) in
Socorro, using an averaging time of 2~s.

Our target source, B3~J2330+3927, was observed in phase-reference
mode.  B3~J2330+3927 and the nearby ($\sim 40'$), International
Celestial Reference Frame (ICRF) source J2333+3901 were alternately
observed through each six-hour long VLBA run.  The observations
consisted of $\sim180$~s scans on B3~J2330+3927 and $\sim100$~s scans
on J2333+3901, plus $\sim20$~s of antenna slew time.  In each
observing run, the total on-target time was $\sim$3.5~h.  J2253+1608
was observed as fringe finder, and was also used to calibrate the
bandpass.

We performed standard a priori gain calibration within {\it AIPS},
using the measured gains and system temperatures of each antenna.  We
fringe-fit and imaged the data for the calibrator J2333+3901 in a
standard manner.  The final image for J2333+3901 was then included as
an input image in a new round of fringe-fitting for J2333+3901.  In
this way, the results obtained for the phase delays and delay-rates
for J2333+3901 were structure-free.  These new values were then
interpolated and applied to B3~J2330+3927 using {\it AIPS} standard
procedures.  These technicalities allowed us to estimate the positions
of components inside B3~J2330+3927 with milliarcsecond astrometric
precision (see Table~\ref{fluxtable}).

To properly image the arcsecond-wide source B3~J2330+3927 with
milliarcsecond resolution, we kept the averaging integration time to
2\,s and used a maximum channel bandwidth in the imaging process of
2\,MHz at each frequency, which results in a maximum degradation of
the peak response for the component furthest away from the phase
center of less than $\sim$4\% and prevents artificial smearing of the
images (e.g., \citealt{bs99}).  We used standard phase
self-calibration techniques within {\it AIPS} to image B3~J2330+3927
at both frequencies.  The image off-source rms was $\sim 60\,\mu$Jy at
both 1.7 and 5.0\,GHz.

\begin{figure}
\begin{center}
\hspace{-1cm}
\psfig{file=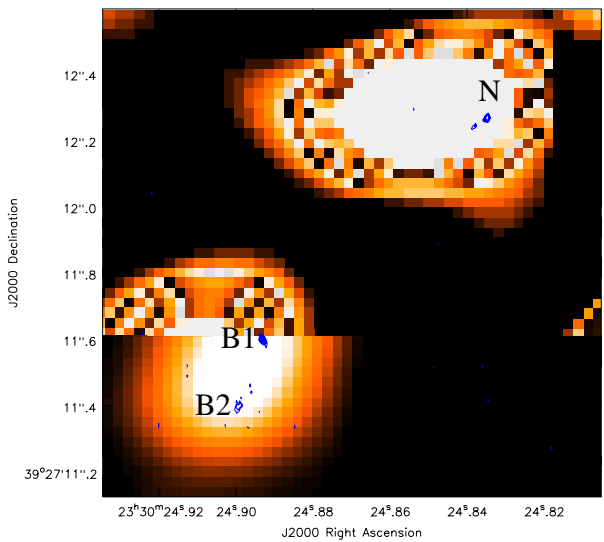,width=8cm}
\caption{Central part of the VLA 8.4\,GHz image showing components N
and B with the VLBA 1.7\,GHz image overlaid as blue contours. Contour
levels are a geometric progression in $\sqrt{2}$, with the first
contour starting at 0.36\,mJy beam$^{-1}$. 
}
\label{fig,vlavlba}
\end{center}
\end{figure}

\begin{figure}
\begin{center}
\hspace{-1cm}
\psfig{file=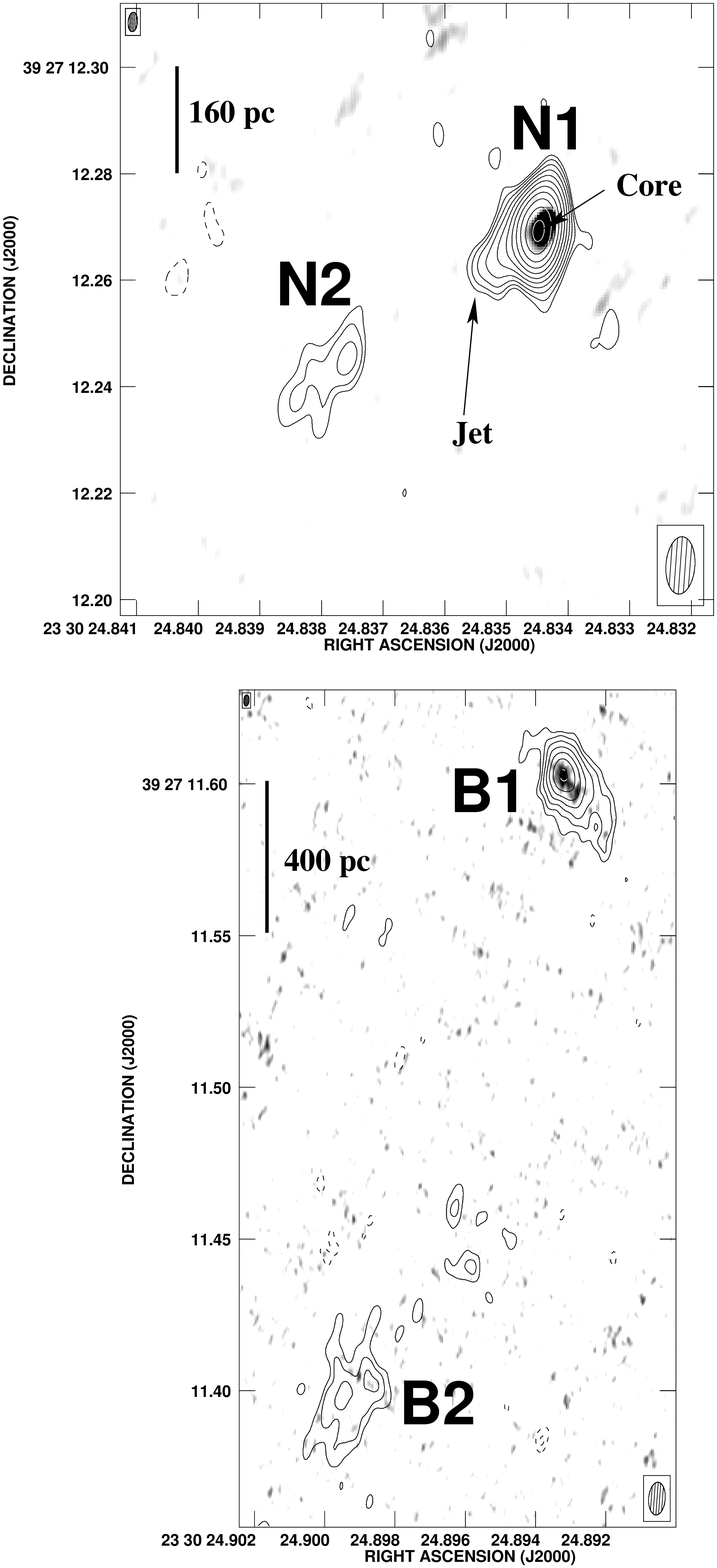,width=8.2cm}
\caption{VLBA 5.0\,GHz images of components~N (top) and B (bottom)
 shown in greyscales with VLBA 1.7\,GHz contours overlaid. Contour
 levels are a geometric progression in $\sqrt{2}$, with the first
 contour starting at 0.1\,mJy beam$^{-1}$. Note the protrusion
 southeast of component N1, which corresponds to the jet emanating
 from the radio core, as confirmed by the detection of N2.  Note that
 the jet bends by $\approx$35\degr between N2 and B1 (see text), and
 that B1 is extended perpendicular to the jet axis.}
\label{fig,vlba}
\end{center}
\end{figure}

\section{Results and Discussion}
\label{results}

Figure~\ref{fig,vla} shows the reanalyzed 8.4~GHz VLA data on
B3~J2330+3927, with the archival 1.4~GHz data overlaid.  In addition
to the three components identified by DB03, we also detect a faint
component on the northwestern side of object~{\em a}.  In the
following, we call these components A, B, C, and N, as identified in
Fig.~\ref{fig,vla} and Table~\ref{fluxtable}.  The newly found
component A, shows for the first time the detection of the counter-jet
in B3~J2330+3927, providing support for the interpretation that the
radio component N, coincident with object~{\em a}, is the radio core.

We used the VLBA to image with mas-resolution the entire VLA region,
encompassing objects A through C.  We detected components N and B
only, both at 1.7 and 5.0~GHz. Figure~\ref{fig,vlavlba} shows the
8.4~GHz map of these components, with the 1.7~GHz VLBA map
overlaid. Figure~\ref{fig,vlba} displays blow-outs of components N and
B to show the VLBA data at their full resolution.  Our 1.7~GHz VLBA
data shows that region N consists of a strong source N1, composed of a
very compact region and a protrusion which likely corresponds to the
jet, and a relatively faint, jet-like feature, N2, at a position angle
of $\approx$125\degr. Component B also splits into two components, B1
and B2, at a position angle of $\approx$160\degr.  Component B1 shows
an elongated structure at a position angle of $\approx$-132\degr, thus
forming an angle of 68\,\degr with respect to the line connecting B1
and B2.  Our 5.0~GHz VLBA data only detect components N1 and B1.
Table~\ref{fluxtable} lists the flux densities and positions for all
these VLBA and VLA components.  Component N1 has a spectral index
$\alpha_{5.0}^{1.7}$=$-$0.2$\pm$0.1 and B1 has
$\alpha_{5.0}^{1.7}$=$-$1.2$\pm$0.1, while components N2 and B2 both
have $\alpha_{5.0}^{1.7}$$\lsim$$-$0.9.  Because component N1 has the
flattest spectral index of all components and is very compact, we
identify it with the radio core of B3~J2330+3927, which is located at
the position of the optical/near-IR type~II AGN and the peak of the
CO(4-3) emission (DB03).  We note that the extrapolated VLBA flux
density at 8.4\,GHz for N1 ($\approx$10.8\,mJy) is very close to the
observed VLA flux, implying that most of the VLA emission comes from
the core.  Indeed, the fraction of the 8.4~GHz core emission, $f_c
\approx 0.50$, is extremely large for a radio galaxy, and even large
for quasar standards \citep{saikia95}.  The extrapolated VLBA flux
density from component B1 at 8.4~GHz is $\approx$1.5\,mJy,
substantially lower than observed, indicating that a significant
amount of flux density comes from a rather extended region.

\setcounter{table}{0}
\begin{table}
\centering
\caption{Gaussian component fits of B3~J2330+3927.\label{fluxtable}}
\begin{tabular} {llllcl} \hline
\multicolumn{1}{l}{$\nu$}      &
\multicolumn{1}{c}{Comp.}  	   &
\multicolumn{1}{c}{$\Delta$RA}  	   &
\multicolumn{1}{c}{$\Delta$DEC} 	   & 
\multicolumn{1}{c}{$S_{\rm peak}$} &
\multicolumn{1}{c}{$S_{\rm tot}$}  
\\ 
\multicolumn{1}{l}{{\scriptsize GHz}}   &
\multicolumn{1}{c}{Name}    &
\multicolumn{1}{c}{{\scriptsize J2000}} &
\multicolumn{1}{c}{{\scriptsize J2000}} &
\multicolumn{1}{c}{{\scriptsize mJy/beam}}        &
\multicolumn{1}{c}{{\scriptsize mJy}}   
\\ 
\hline 
1.7$^\dagger$ & N1 &   24\fs 8345 &  12\farcs 269 &10.45 &  $13.47 \pm 0.68$  \\ 
    & N2 &   24\fs 8381 &  12\farcs 242 & 0.71 &  $ 1.89 \pm 0.11$ \\
    & B1 &   24\fs 8932 &  11\farcs 603 & 2.45 &  $ 9.89 \pm 0.50$  \\  
    & B2 &   24\fs 8997 &  11\farcs 397 & 0.71 &  $ 5.53 \pm 0.28$ \\ 
5.0$^\dagger$ & N1 &   24\fs 8345 &  12\farcs 269 & 6.70 &  $11.38 \pm 0.57$ \\ 
    & B1 &   24\fs 8932 &  11\farcs 602 & 0.35 &  $ 2.77 \pm 0.15$ \\ 
8.4$^\ddagger$ & A  &   24\fs 780  &  13\farcs 12  & 0.31 &  $ 0.35 \pm 0.05$ \\ 
    & N  &   24\fs 837  &  12\farcs 31  &10.02 &  $10.80 \pm 0.54$ \\ 
    & B  &   24\fs 900  &  11\farcs 53  & 3.74 &  $ 6.28 \pm 0.31$ \\ 
    & C  &   24\fs 923  &  10\farcs 63  & 2.55 &  $ 3.98 \pm 0.18$ \\ 
\hline 
\end{tabular}
\begin{list}{}{}
\item[] { \rm The component names are marked in Figs.~\ref{fig,vla}, 
\ref{fig,vlavlba}, and \ref{fig,vlba}; the reference coordinates are
23\h 30\m 0\fs 0 for RA, and 39\degs 27\amin 0\farcs 0 for DEC.
 The standard errors in
the quoted VLBA positions are $\approx$1~mas in each
coordinate, while those for the VLA positions are estimated to be
$\approx$30~mas.
The standard errors in the flux density estimates are the result of
adding in quadrature the image off-source rms 
and a 5\% of the peak flux density at each frequency, to account for
inaccuracies of the system calibration. 
$^\dagger$ From VLBA observations at 1.7 and 5.0~GHz.
$^\ddagger$ From VLA observations at 8.4~GHz. 
}
\end{list}
\end{table}

With the new information yielded by our VLBA observations, we can now
describe a consistent picture of the radio structure of B3~J2330+3927:
the source radio emission is core-dominated, and its morphology very
asymmetric, with a lobe distance ratio $Q = (N1-C)/(N1-A) \approx$1.93
(see Table~\ref{fluxtable}), and a lobe flux density ratio $R = S_{\rm
C}/S_{\rm A}\gsim$11 (from the 8.4~GHz VLA observations).  We can
exclude significant radio emission beyond component A, because our
wide-field 1.4~GHz VLA map (1\degr$\times$1\degr) is very deep, and no
other components are seen in the WENSS \citep{rengelink97} or NVSS
\citep{condon98}.  The change in position angle between the jet-like
feature emanating from the core, N1-jet and N2, and the line
connecting components B1 and B2, indicates that the radio jet is
deflected between components N and B. Jet bending can be naturally
explained by the interaction of the jet with its ambient medium
\citep[\eg][]{saxton05} or, alternatively, as the precession of the
jet in a binary black hole system (e.g., \citealt[]{hummel88}).
Interestingly, no increase in the Ly-$\alpha$ emission is seen in the
2D spectrum (Fig.~4 of DB03), in contrast to similarly deflected radio
jets in other HzRGs, showing bright Ly-$\alpha$ emission at these
bendings \citep[e.g.][]{vanojik96}.

The values of $R$ and $Q$ obtained for B3~J2330+3927 are more common
of quasars than radio galaxies, although statistical analyses by
\citet{mccarthy91} did not show these differences to be highly
significant.  However, \citet{best95} show convincing evidence that
quasars have more asymmetric radio morphologies than radio galaxies.
This finding is consistent with one of the prime expectations of the
unified model, namely that quasars and radio galaxies are
intrinsically similar, but quasars are observed when the radio jet
axis of the source is within 45\,\degr\ of the line of sight
\citep{barthel89}.  In addition, quasars also display larger $f_c$
values than radio galaxies \citep{saikia95}.  Therefore, based on its
radio morphology, B3~J2330+3927 would be classified as a type~I
AGN. However, the optical and near-IR Keck spectra (Fig.~5 of DB03)
clearly show only narrow emission lines with relatively strong \HeII\
and very faint \Hbeta\ typical of type~II AGN.  Moreover, the optical
and near-IR continuum emission is weak ($F[\lambda_{\rm rest}=1500{\rm
\AA}]$\,$\approx$\,$5\,\times\,10^{-19}$\,\ergs\,
cm$^{-2}$\,\AA$^{-1}$) and relatively red ($F_{\lambda} \propto
\lambda^{2.7}$), as is common in type~II AGN.

We now examine three possibilities to explain the observed asymmetries
of the radio structure of B3~J2330+3927: (i) environmental effects,
(ii) relativistic beaming effects, and (iii) intrinsic asymmetries. To
date, almost all cases of asymmetry in radio galaxies have been
explained by environmental effects. In the most detailed study to
date, \citet{mccarthy91} find that for radio-loud type~II AGN, the
radio lobe closest to the core always lies on the same side of the
nucleus as the high surface brightness optical line emission. This is
contrary to what is seen in B3~J2330+3927. Fig.~4 of DB03 shows that
the \Lya emission extends from the radio core N to the southern radio
lobe~C, which is almost twice further away from the core than the
faint radio lobe~A. One could argue that the \Lya\ emission towards
lobe~A is quenched by dust, which is revealed by its strong far-IR
emission \nocite{stevens03}(Stevens et al. 2003, DB03). However, the
spatial profiles of the non-resonant \OII\ and \OIII\ lines in the
Keck/NIRC spectra of DB03 show a similar shape, with no emission
towards lobe~A, indicating that no substantial emission line flux is
missing in the \Lya\ profile. Environmental effects thus do not seem
to play a major role in the radio morphology of B3~J2330+3927.

Turning to relativistic beaming effects, we note that jet speed,
$\beta$, viewing angle, $\phi$, and arm-length ratio, $Q$, are related
by $\beta\,cos\,\phi = (Q-1)/(Q+1)$ (e.g., \citealt{best95}).  The
fact that we observe a clear type~II AGN indicates that the viewing
angle with respect to the line of sight should be
$\phi\,>$45\degr. However, the fraction of the emission from the
nuclear component, $f_c$=0.50 at 8.4~GHz, is very large for a radio
galaxy, and suggests the viewing angle must be then close to the
$\phi$=45\degr\ limit.  If this is the case, the obtained arm-length
ratio $Q\approx$1.9 requires then a jet speed of $\beta \approx 0.45$.
Now, the observed ratio of the flux densities of components C and A is
$R\gsim$11. If components A and C are optically-thin, isotropically
emitting jets, their flux ratio should then be, within the
relativistic beaming scenario (e.g., \citealt{mr99} and references
therein), equal to
$\big(\frac{1+\beta\,cos\theta}{1-\beta\,cos\theta}\big)^{k-\alpha}$,
where $k$=2 for a continuous jet, and $k$=3 for the case of discrete
condensations.  Assuming that components A and C have an spectral
index of $\alpha$=$-$0.9, the flux ratio for a continuous jet would be
$\lsim$4, whereas for discrete condensations would be $\lsim$6.  Our
observed value of $R\gsim$11 is therefore difficult to explain within
the standard relativistic beaming framework.

The most likely explanation of the observed asymmetry thus seems to be
an intrinsic difference in the radio jets. Examples of such radio
galaxies are very rare in the literature, and have only been reported
at low redshifts, for example in the $z$=0.290 radio galaxy 4C~63.07
\citep{saikia96}.  Sources like B3~J2330+3927 and 4C~63.07 are
difficult to reconcile with predictions from the standard unified
model for quasars and radio galaxies \citep{barthel89}.

We estimated the energetics and magnetic fields in the regions N and B
of B3~J2330+3927, using our VLBA images.  We measured the angular
sizes of components N1 and B1, and used their flux densities and
spectral indices between 1.7 and 5.0~GHz to characterize them using
standard prescriptions (e.g., \citealt{miley80}).  We assumed spectral
cutoff frequencies of $\nu_1$=0.01~GHz and $\nu_2$=100~GHz, a ratio of
heavy particles to electrons $k$=100, and a filling factor of the
emitting regions of 1.  N1, the core, is characterized by a
synchrotron radio luminosity of $L_{\rm R} \sim 3\EE{45}$\ergs, an
average magnetic field $B \sim 7$~mG, a minimum energy density $u_{\rm
me} \sim 5\EE{-6}$\erg \cm3, and a (minimum) total energy $E_{\rm tot}
\sim 7\EE{56}$~erg.  The radio region B1 displays similar values: $B
\sim 14$~mG, $u_{\rm rel} \sim 2\EE{-5}$\erg \cm3, $E_{\rm tot} \sim
8\EE{57}$~erg, $L_{\rm R} \sim 8\EE{44}$\ergs.  Our estimates of the
magnetic field in those components are in good agreement with magnetic
field estimates for other HzRGs (e.g., \citealt{cai02}), and
significantly higher than estimates found in typical nearby powerful
radio galaxies (e.g., $B \sim 3\EE{-4}$ for Cygnus A; \citealt{cb96}).
If there is no energy resupply to the above radio source components, a
radiative source component lifetime can be estimated as $\tau = E_{\rm
rel}/L_{\rm R}$, where $E_{\rm rel} = 7/4\,(1+k)^{-1}\,E_{\rm tot}$
(\citealt{pacholczyk70}).  This leads to $\tau \sim 2\EE{4}$~yr and
$\tau \sim 7\EE{5}$~yr for N1 and B1, respectively.  Since the average
age of the relativistic electrons, $t_r$, in N1 and B1 is in both
cases $\sim$600~yr, i.e., $t_r \ll \tau$, it follows that a continuous
supply of newly accelerated electrons is required to explain the
observed radio emission in those regions.

\section{Summary}
\label{summary}

The main results of our VLBA and VLA analysis of B3~J2330+3927 can be
summarized as follows:

\begin{enumerate}
\item
We identify component N1, a rather compact, flat spectrum ($\alpha
\approx -0.2$) with the core. It is located at the position of the
optical/near-IR type~II AGN and the peak of the CO(4-3) emission.

\item
The 1.4 and 8.4\,GHz VLA images reveal a previously undetected, very
faint counterjet. The 8.4~GHz VLA image shows that B3~J2330+3927 has a
radio lobe flux density ratio $R\gsim$11, a radio lobe asymmetry ratio
$Q\approx$1.9, and a fraction of core emission $f_c \approx 0.50$.
These values are much more compatible to quasars, and not expected to
occur in radio galaxies.  However, B3~J2330+3927 is a bona-fide
type~II AGN by virtue of its optical and near-IR spectra.

\item
The fact that almost all the \Lya\ emission is located at the side
where the radio lobe is further away from the core is contrary to what
has been observed in other radio galaxies. This argues against
environmental effects as the origin of the radio source
asymmetry. Relativistic beaming effects also cannot explain the
observed flux density asymmetry, implying an intrinsic asymmetry of
the radio jets. This is the first example of such a source at high
redshift, and seems to be difficult to reconcile with the unified
model.

\item 
We estimated the energetics and magnetic field 
of the innermost regions of B3~J2330+3927. We find values
of the order of $B \sim$7-15~mG for N1 and B1, values that  are 
significantly higher than estimates found in typical nearby powerful
radio galaxies (e.g., $B \sim 0.3$~mG for Cygnus A).
From the estimated radiative lifetimes for those regions, 
we conclude that a continuous supply of newly
accelerated electrons is required to explain the observed radio
emission.

\end{enumerate}

\section*{Acknowledgments}
We thank R. Ivison for providing the calibrated 1.4\,GHz VLA data,
R. Laing for stimulating discussions, and an anonymous referee for a
prompt and useful report. This research has been partially funded by
grant AYA2001-2147-C02-01 of the Spanish Ministerio de Ciencia y
Tecnolog\'{\i}a.  MAPT is supported by the programme Ram\'on y Cajal
of the Spanish Ministerio de Educaci\'on y Ciencia.  NRAO is a
facility of the National Science Foundation operated under cooperative
agreement by Associated Universities, Inc.  This research has made use
of NASA's ADS.

\label{lastpage}
\end{document}